# Field-Selective Adsorption of Saccharin on Nickel: Mechanistic DFT Insights into Solvation, Protonation, and Coating Morphology

Aylar G. M. Ghashghaei [1], Mahboubeh Khorrami[2] , Mohammad Ebrahim Bahrololoom[3]

[1] Department of Materials Science and Engineering, Shiraz University, Shiraz, Iran

Email: ailarghashghaee@yahoo.com

[2] Department of Chemistry, Michigan State University, East Lansing, MI 48824, USA

Email: khorram4@msu.edu

[3] Professor, Department of Materials Science and Engineering, Shiraz University, Shiraz, Iran

Email: bahrolom@shirazu.ac.ir

## Abstract
The molecular mechanisms by which organic additives such as saccharin control microstructure in nickel Electrodeposition remain inadequately understood, particularly the role of the intense interfacial electric field. This study employs density functional Theory (DFT) calculations to elucidate the field dependent adsorption behavior of neutral saccharin and its deprotonated anion (saccharinate) on nickel. By employing the B3LYP functional and implicit solvent models, the field dependent adsorption energetics, Frontier orbitals and electrostatic potentials are calculated on a nickel surface. Key findings reveal that while saccharinate dominates in bulk plating baths, its strong solvation shell impedes surface adsorption. In contrast, neutral saccharin exhibits energetically favorable adsorption via sulfonyl oxygen or aromatic π-face interactions, with specific orientations further stabilized by the interfacial field.This selective adsorption at growth sites rationalizes saccharin's role in inhibiting rapid crystallization, promoting grain refinement, and producing bright, level deposits.The results directly link field-modulated molecular stereochemistry to macroscopic coating properties,providing a mechanistic foundation for the rational design of electroplating additives beyond empirical approaches .

## Introduction
Nickel electrodeposition is a mature yet continually evolving electrochemical technology for producing high quality metal coatings. The process involves the cathodic reduction of nickel ions from aqueous electrolyte to form a metallic film on the substrate. This technique is widely adopted because it combines cost-effectiveness, operational simplicity, and scalability with high current efficiency and effective regulation over coating thickness and morphology.Moreover, electroplated nickel provides multiple benefits including protection against corrosion and wear, aesthetic enhancement through bright and uniform surfaces, and various technical applications such as solderable base layers, barriers against diffusion and magnetic or catalytic surfaces. These diverse applications explain its extensive use in industries ranging from automotive and aerospace to electronics, manufacturing tools, and consumer products.

The quality of the final coating is heavily influenced by the chemical makeup of the solution, process conditions and the inclusion of organic compounds. Traditional Watts solutions (containing $NiSO_4$–$NiCl_2$–$H_3BO_3$) without organic compounds usually produce soft and non reflective coatings. Therefore, contemporary plating solutions incorporate small amounts of additives to modify the coating's texture, mechanical characteristics, and visual appearance. However, despite widespread implementation, the molecular level mechanisms through which these additives alter nucleation and growth remain only partially understood. it is known that additives can act as wetting, leveling, brightening or buffering agents[1].Among these additives, saccharin (1,2-benzisothiazol-3(2H)-one-1,1-dioxide) has been one of the widely applied agents in nickel electroplating for more than a century. Saccharin exists in three main forms (table 1), namely neutral saccharin (acid form), sodium saccharin, and calcium saccharin. Among these, sodium saccharin is the most commonly used in electroplating baths due to its high water solubility[2]. Numerous studies have demonstrated that saccharin promotes the formation of fine-grained, compact deposits, thereby enhancing surface brightness, hardness, and wear resistance[3-5]. Another effect is the alteration of residual stress within the nickel layer and converting internal stress from tensile to compressive, which is particularly advantageous for preventing crack formation and delamination in thick coatings.

*Table 1 Saccharin forms and corresponding structures[2]*

| Forms of saccharin | IUPAC Name | Structure |
|---|---|---|
| **Neutral saccharin** | 1,2-Benzisothiazolin-3-one | 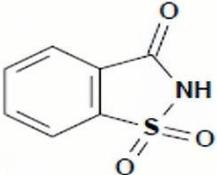 |
| **Sodium saccharin** | 1,2-Benzisothiazolin-3-one, 1,1-dioxide, sodium salt | 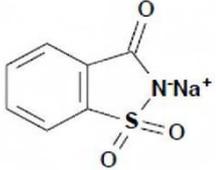 |
| **Calcium Saccharin** | 1,2-Benzisothiazolin-3-one, 1,1-dioxide, calcium salt | 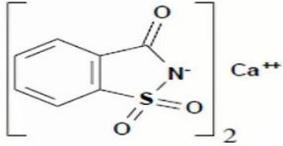 |

Proposed explanations for these effects include reduced nickel mobility, charge-transfer blocking, or inhibition of growth at specific crystal planes[6].Through real-time scanning probe microscopy studies, researchers have directly observed saccharin's impact on nickel film development, showing that while it has minimal effect on surface movement at low overpotentials, it does interfere with three-dimensional crystal formation, changing the coating's surface characteristics and initially increasing roughness before smoothing in thicker films[7].A recent kinetic study on Ni-Co electrodeposition using electrochemical impedance spectroscopy (EIS) showed that saccharin's behavior is governed by a dynamic competition between inhibition and site selective separation effects.At low concentrations, it promotes site separation of metal reduction and hydrogen evolution, resulting in rough, impure deposits. At higher concentrations, inhibition dominates, suppressing hydrogen evolution, stabilizing local pH, and minimizing oxide impurities, thereby producing compact, high-purity films with refined grains and increased hardness[8].

Various In-situ spectroscopic studies (surface enhanced Raman scattering, SERS) confirm that saccharinate anions bond via the carbonyl oxygen and imide nitrogen to transition-metal surfaces, with the aromatic ring oriented toward the electrolyte. For example, according to Kwon and Gewirth, in the binding of the saccharinates to the surface of the nickel electrode at a potential above the potential of the zero metal charge, the binding occurred through the coordination of the oxygen atoms, while at a higher negative potential, a bis-ligated binding occurred through the coordination of the oxygen and nitrogen atoms. This observation indicates that the electrode potential, and thus the electric field present at the interface, can control the binding position and orientation of saccharin[9]. Moreover, it has been observed that for direct current, pulsed current, and pulse-reverse current electrodeposition, the continuous adsorption of saccharin at the growth centers blocks the growth of the grain vertically, but for pulse-reverse currents, the desorption of saccharin during the anodic pulse leads to the enhanced growth of the grain vertically[10]. These results indicate that saccharin primarily influences 3D growth, though the precise mechanism, possibly orientation dependent or geometric, remains to be clarified.

In the process of nickel electrodeposition, the electrical double layer (EDL) creates an organized and dynamic interface that, in the case of the cathode, influences the processes of electric charge transfer and adsorption. With the application of an electric negative potential to the cathode surface, the excess electric charges are compensated for by the organized solvated cations, and the multilayered region comprises the inner helmholtz plane (IHP), where the oriented water molecules and specifically adsorbable species are present, and the outer Helmholtz plane (OHP), followed by the diffuse layer. The intense electric fields in this compact region which often exceeding $10^9$–$10^{10}$ V·m$^{-1}$, directly influence water orientation, local dielectric properties, and ion solvation/desolvation behavior. Modern EDL descriptions, including the Gouy–Chapman–Stern and Grahame models, emphasize their dynamic nature, continuously reorganizing with changes in potential, electrolyte composition, and additive adsorption. As a result, Ni$^{2+}$ reduction occurs within this nanometer scale environment where ion migration, partial desolvation and electron transfer are tightly coupled. Organic additives, such as saccharin, cause significant perturbations to the electric double layer due to the competition for the native IHP adsorption sites, the influence on the distribution of electric charges, the differential capacitance, and the electric field. As such, the electric double layer becomes an active interface that can be manipulated for controlling the processes of nucleation, inhibition, grain refinement, and the microstructure development of the nickel coating deposits[11, 12].

In order to manage the deposit microstructure, brightness, and strength rationally from the molecular level interaction point of view, it is essential to understand the interaction behaviors of the additives. Although saccharin and various other molecular additives are widely applied, there are uncertainties related to their interaction behaviors. Computational approaches provide a promising route to address these gaps. On the contrary, based on the fundamental laws of physics, DFT calculations provide direct and reliable predictions and enables us to simulate numerous physiochemical/electronic parameters of materials in vacuum and in solvation at the atomic scale[13]. In particular, DFT calculations can provide reliable predictions regarding the optimal adsorption geometry, adsorption energy, electronic charges transferred, and the density of states and frontier orbitals, which are often difficult to experimentally investigate[14]. Crucially for electrochemical systems, DFT can explicitly include bias potential or applied electric field, enabling systematic evaluation of how field direction and magnitude modify adsorption orientation, binding strength and induced dipoles at the interface[15]. In such a manner, the integration of atomistic and field driven information coupled with experimental facts creates a solid foundation for the explanation of adsorption-driven processes involving the influence of additives on the nucleation, growth, and properties of the electrodeposited nickel.

In the present study, DFT calculations are utilized for explaining the molecular processes occurring during the electrodeposition of nickel that are affected by the saccharin molecule and its deprotonated form, the saccharinate ion. This focus on the interaction between the electric field and the adsorption stereochemistry of molecules towards nickel substrate offers clearer routes to microstructure engineering, in contrast to the trial and error approach commonly used in additive design. These calculations elucidate the favorable geometries and binding energetics for adsorption at representative low-coordinate sites, as well as the attendant charge transfer. Frontier molecular orbital analysis reveals possible orbital-level interactions between the additive and metal states, while Fukui-function analyses and molecular electrostatic potential (MEP) maps identify nucleophilic and electrophilic sites that govern directional chemisorption. Through the integration of calculated adsorption patterns, binding energetics, and fingerprints into the phenomenology of nickel plating, this theoretical study provides mechanistic insights and guidance for process optimization.

## Materials and Methods
### Computational method

All electronic structure calculations were conducted using the Density Functional Theory approach with the Gaussian 16 package. The B3LYP (Becke three-parameter Lee–Yang–Parr) hybrid functional was employed due to its well-established reliability. A 6-311++G(d,p) basis set was used for all atoms, providing a flexible valence triple-ζ description enhanced by diffuse and polarization functions to accurately model species and bonding characteristics[14]. To model additive-Ni interactions, a single Ni atom was paired with each saccharin molecule and its deprotonated saccharinate form in various orientations. Implicit solvation was included via the Polarizable Continuum Model (PCM) to approximate aqueous conditions during geometry optimizations and energy evaluations[16]. In each case the Ni-additive complex was geometry optimized to find the lowest energy configuration. This procedure effectively samples different approach geometries of the additive with respect to the Ni center to capture possible coordination modes or binding sites. Furthermore, external electric fields with different magnitudes were applied as a homogeneous field along the x-direction. For each optimized Ni-additive complex, the adsorption (binding) energy ΔE was calculated using the formula:

$\Delta E = E_{adsorbed\ system} - [E_{Ni} + E_{additive}]$     eq (1)

Here $E_{adsorbed\ system}$ is the total electronic energy of the Ni-additive complex, $E_{Ni}$ is the energy of the isolated Ni atom in its optimized geometry , and $E_{additive}$ is the energy of the free additive molecule prior to adsorption. Negative ΔE indicates energy is released during the formation of complexes, and it quantifies the strength of the binding of Ni and the additive.

The ability of the B3LYP function to predict structure and energy has been proven and it is known to follow the standard form of the Kohn-Sham equation. This method allows for a detailed examination of additive adsorption on Ni at the molecular level. In Kohn–Sham DFT, the total electronic energy E[ρ] of a system can be expressed as a sum of kinetic, electrostatic, and exchange correlation contributions:

$E[\rho] = T_s[\rho] + E_e[\rho] + E_{xc}[\rho]$     eq (2)

where $T_s$ is the non interacting kinetic energy of the electrons, $E_e$ includes all classical Coulomb terms (nuclear-electron attraction plus electron-electron repulsion), and $E_{xc}$ is the exchange correlation energy. In practice, the exchange correlation term $E_{xc}$ accounts for the quantum mechanical corrections beyond the classical Coulomb picture[5, 17, 18].

## Interfacial Electric Fields

External electric fields were used to simulate interfacial fields under which an adsorbate is exposed on the IHP during electrodeposition. The approach used here facilitates a thorough analysis of the effects of the orientation as well as the strength of the interfacial electric field on the orientation and bonding characteristics of the additive on a Ni surface. For a nominal potential difference V applied across a gap of width d (the Ni-adsorbate separation), the uniform electric field in SI units is:

$$E_{SI} = \frac{V}{d} \qquad \text{eq (3)}$$

To convert to atomic units (a.u.) used in the quantum-chemistry input, we divide by the atomic unit field $E_0 = 5.14220652 \times 10^{11}$ V m$^{-1}$:

$$E(a.u.) = \frac{V/d}{5.14220652 \times 10^{11}} \qquad \text{eq (4)}$$

The applied electric fields (0.005–0.02 a.u.) correspond to approximately $2.6 \times 10^9$–$1.0 \times 10^{10}$ V m$^{-1}$ (≈ 0.26–1.03 V Å$^{-1}$), representing the strong local field expected within nanometer scale double layers, particularly across the IHP and near charged surface asperities. This range captures realistic interfacial conditions, equivalent to potential drops of about 1 V across separations of 1–2 Å. An external electric field was applied using the Field=Read keyword in the Gaussian input. The magnitude and direction of the electric field were then specified at the end of the Cartesian coordinate section.The field strength was calibrated to represent specific potential drop, for example value of -0.0130 a.u. corresponds to 1 V across a 1.5 Å gap. It should be noted that the homogeneous external field used in this PCM based DFT calculations is an approximation to the nonuniform fields present within the electrochemical double layer[19, 20].

## Results and discussion

DFT computations were used to analyze the relationships between geometric/electronic properties of individual saccharin molecules and their ability to adsorb on surfaces. In Figure 1, the atomic structure and frontier orbitals of the saccharins are shown, which characterise the corresponding electronic properties and reactivity. Molecular reactivity has been inferred to a large extent on the basis of frontier orbitals, characterised by the HOMO (highest occupied molecular orbital), LUMO (lowest unoccupied molecular orbital), and the corresponding energy gap (ΔE). The localized orbitals match the chemical character of the molecule, which has an aromatic π-system and a sulfonamide moiety (–SO$_2$NH), contributing nitrogen and oxygen lone pairs. These can act as relatively strong σ-donor sites (lone electron pairs on N/O atoms) and weak π-donors (aromatic π-electrons). This makes saccharin capable of reorientation and polarization due to strong electric fields existing in the electrochemical double layer[21]. In the context of saccharin as an additive in nickel electrodeposition, the adsorption strength of the molecule on the cathode plays an important role in modulating the cathodic polarization of nickel. The molecule binds to the nickel by donor-acceptor bonding, which is determined by the frontier orbitals of the molecule. In fact, it is the HOMO that donates electron density to the empty 3d orbitals of Ni (Ni: [Ar] 3d8 4s2), and it is the LUMO that accepts electron density donated by the metal. Graphical representation of the orbitals is helpful to understand the distribution of electron density and the bonding between molecules and the metal.According to frontier-orbital theory, a smaller HOMO-LUMO gap (ΔE) favors electron transfer because it brings the energy levels of the donor and acceptor closer, allowing easier interaction with the metal. Thus, a smaller ΔE is associated with stronger adsorption and, consequently, more interaction between the

Ni d-orbitals and the saccharin S-O region or π-system of the aromatic ring, which is important during the molecule's approach to the nickel surface[22].

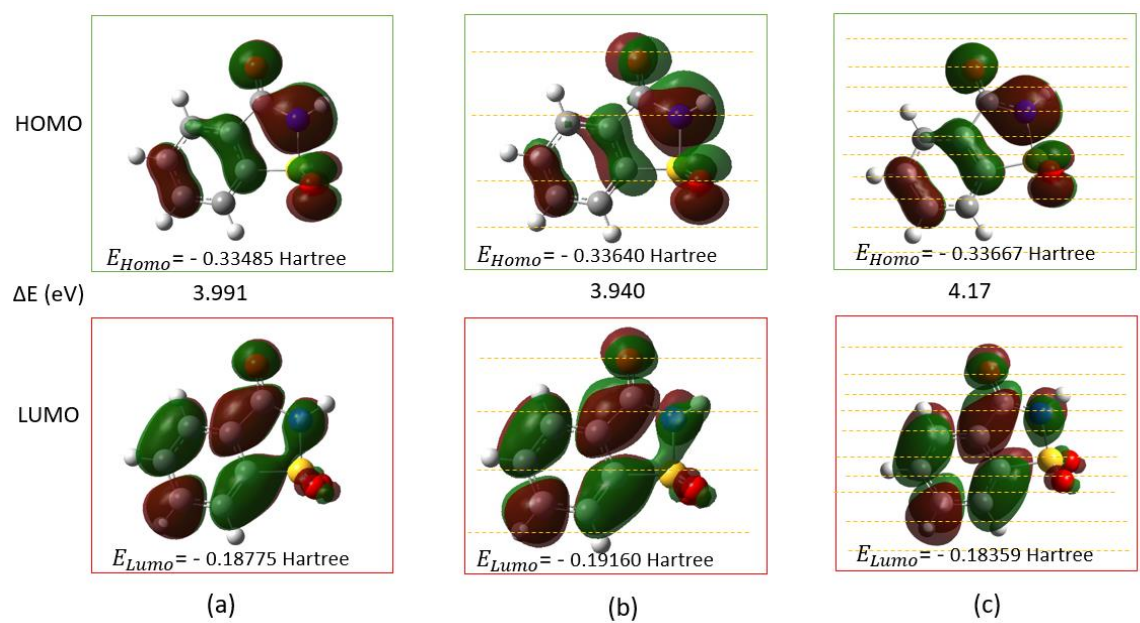

Figure 1 : HOMO and LUMO of saccharin in the gas phase calculated by DFT under three external electric-field conditions: (a) no field, (b) 1 V, and (c) 2 V. Green and red lobes represent opposite orbital phases. The HOMO-LUMO gaps are represented by ΔE and are 3.991 eV (a), 3.940 eV (b), and 4.17 eV (c), respectively.

Application of an external field produces small state dependent changes in the frontier orbitals of saccharin that are relevant to its interaction with metal surface. In the field-free case (Fig. 1a) the HOMO and LUMO are largely delocalized over the aromatic/sulfonyl backbone, yielding a relatively large ΔE of 3.991 eV. An external field of 1 V (Fig. 1b) causes a very small electron density redistribution and a relatively small difference in the Kohn-Sham gap (around 3.94 eV), suggesting an early stage polarization process but without a significant donor-acceptor gap transformation. In the case of the largest external field tested (2V, Fig. 1c), the HOMO and LUMO orbitals show stronger separation and localized distribution on separated fragments of the molecule. Importantly, for the neutral saccharin shown here, the ΔE does not collapse at 2 V which means the field reorganizes orbital density without necessarily making the molecule universally easier to oxidize or reduce. Thus, field-driven orbital localization primarily changes the site selectivity of frontier interactions rather than producing a simple overall increase in chemical softness.

Mechanistic implications for nickel electrodeposition follow from these two effects. First, localization under external fields creates more distinct electrophilic and nucleophilic centers. This means regions of enhanced HOMO density will preferentially donate electron density to coordinatively unsaturated Ni atoms while LUMO-rich sites will be better positioned to accept back donation or stabilize local positive charge. Practically, the sulfonyl oxygens and the ring nitrogen (high HOMO density in many conformers) act as primary σ-donor coordination sites, whereas the aromatic π-system can contribute weaker π-type interactions and additional overlap with Ni d-orbital. Second, the precise energetic alignment of these localized orbitals with Ni d-states determines whether adsorption is dominated by donation, back donation, or a mixed covalent/ionic character. Consequently, external applied surface potentials or local double layer fields are expected to tune not only how strongly saccharin binds to a Ni surface but also influence which functional group and stereochemistry of saccharin are involved in binding. This, in turn, would affect various subsequent processes, like rates of local electron transfer, nucleation, and microscopic structure of the deposited matter, which would relate to grain size, texture, and rate of formation of zerovalent deposits.

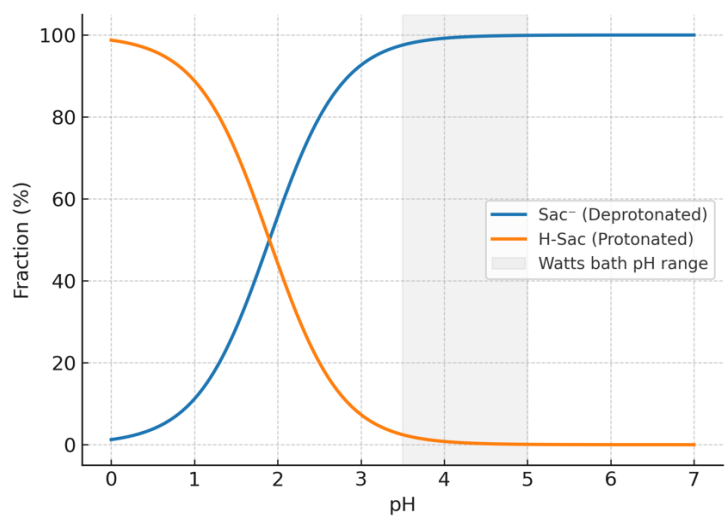

*Figure 2 Speciation of saccharin as a function of pH (pK$_a$ ≈ 1.94). The blue curve denotes the deprotonated saccharinate anion (A⁻) and the orange curve the protonated (neutral) saccharin (HA). The shaded band indicates the typical pH range of a Watts nickel bath (≈3.5–4.5).*

As shown in Figure 2, the speciation of saccharin is strongly governed by the solution pH relative to its acid dissociation constant (pK$_a$ ≈ 1.94). The plot shows that the deprotonated saccharinate anion (Sac⁻) becomes dominant around pH ≈ 2, while the fraction of the protonated H-Sac species decreases. Even with the upper reported pK$_a$ value of 2.3 considered, over 95 % of saccharin in a Watts nickel bath (≈ 3.5- 4.5) is in its anionic form within its normal pH range[2]. These results, obtained from the Henderson–Hasselbalch relation, confirm that under plating conditions saccharin is essentially fully deprotonated. Therefore, computational and mechanistic studies employing only the neutral molecule may not accurately represent the actual interfacial chemistry under plating conditions. The reactive environment at the electrode interface is fundamentally distinct from the bulk solution. For example, in the vicinity of the cathode, the electrochemical double layer generates extreme gradients in electrostatic potential, ion concentration, and local proton activity. Research on organic additives for nickel and copper electrodeposition show that species with small bulk concentrations can reach appreciable surface coverage. This is driven by specific adsorption, field-induced orientation, and shifts in local protonation deprotonation equilibria away from bulk values[9, 23]. Though saccharine is mostly found in its deprotonated anionic form in the bath, the electrode potential significantly changes the local acid-base equilibrium. The strong electric field at the negatively charged cathode can stabilize transiently protonated or neutral forms and change adsorption equilibria toward species that interact better with the metal surface[24]. Therefore, in this study, to ensure a comprehensive understanding, calculations were performed for both the neutral and anionic forms, allowing comparison of their adsorption behavior and electronic characteristics on the nickel.

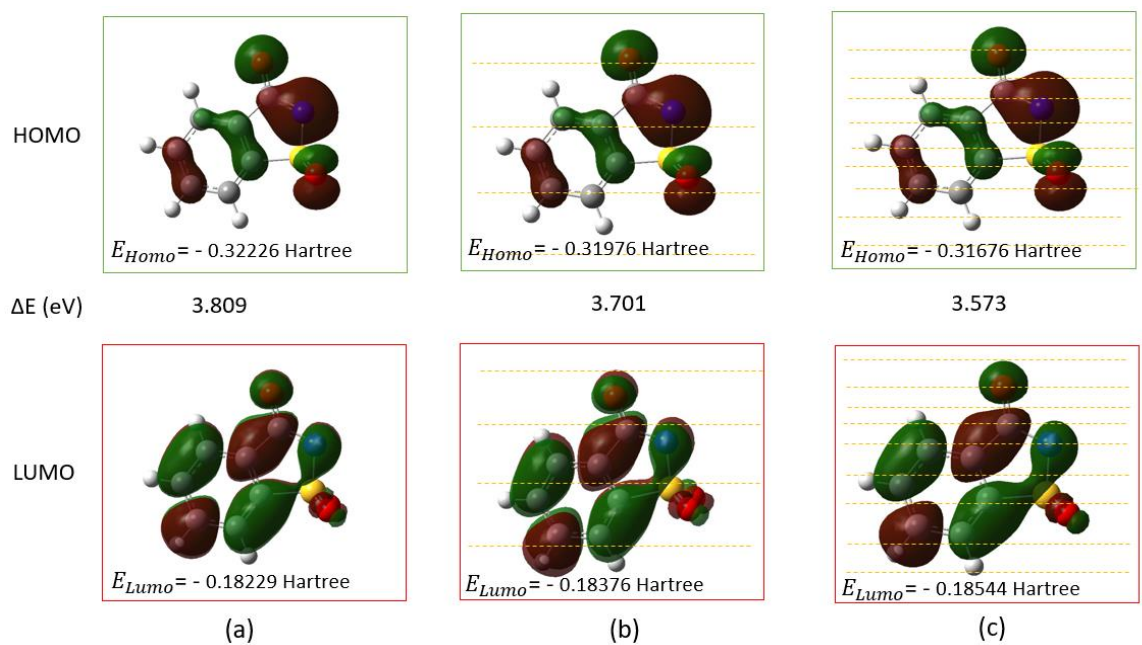

*Figure 3 Frontier molecular orbitals (HOMO, top row; LUMO, bottom row) of the saccharinate anion (deprotonated saccharin) in the gas phase calculated by DFT under three external electric-field conditions: (a) no field, (b) 1 V, and (c) 2 V. Green and red lobes denote opposite orbital phases. Calculated HOMO–LUMO gaps (ΔE) are shown below each column: 3.809 eV (a), 3.701 eV (b) and 3.573 eV (c).*

According to Fig 3, the applied electric field systematically alters the electronic structure of the saccharin anion. The HOMO energy increases slightly from −0.32226 Hartree (0 V) to −0.31676 Hartree (2 V), while the LUMO energy shifts from −0.18229 Hartree (0 V) to −0.18544 Hartree (2 V). Consequently, the Kohn–Sham gap narrows progressively from 3.809 eV (0 V) to 3.701 eV (1 V) and further to 3.573 eV (2 V). This reduction in the HOMO–LUMO gap indicates enhanced electronic softness and polarizability of the anion under an electric field.

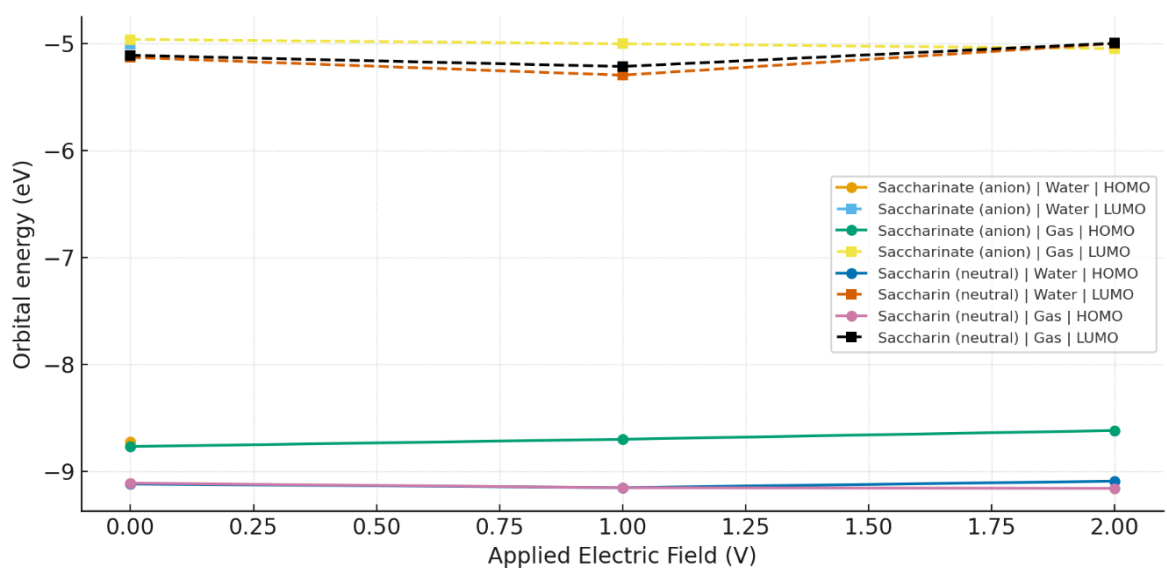

*Figure 4 Field-dependent Frontier Orbital Energetics. Variation of HOMO and LUMO energies for neutral saccharin and the saccharinate anion as a function of applied electric field, computed in gas phase and with implicit aqueous solvation.*

The field dependent evolution of the frontier molecular orbitals (Figure 4) quantifies the electronic response of both saccharin species. The saccharinate anion responds more to the electric field than the neutral molecule, reflected in the greater degree of the HOMO-LUMO energy gap contraction. This systematic reduction in ΔE , particularly under solvated conditions, enhances the anion's chemical softness and polarizability. As a result , the interfacial electric field at the cathode might actively tunes the saccharinate anion into a more reactive state, thereby strengthening its coordinative adsorption to the nickel via improved donor-acceptor coupling.

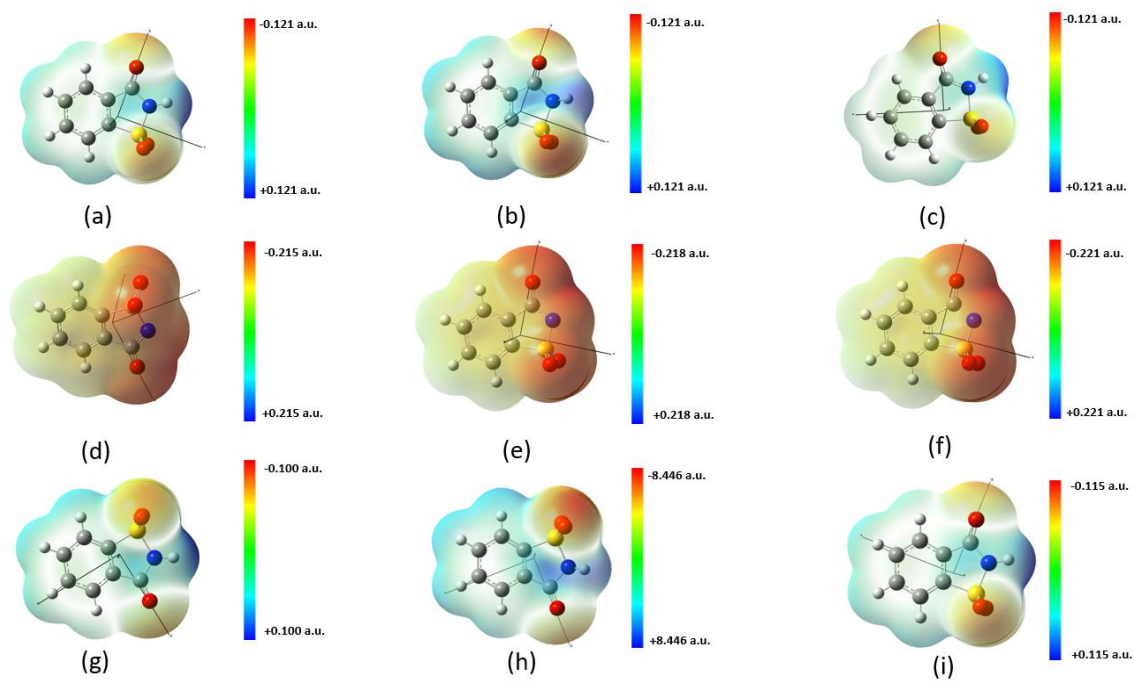

Figure 5 Molecular electrostatic potential mapped onto the electron density isosurfaces for saccharin (neutral) and saccharinate (deprotonated) under three conditions: gas phase without applied field (panels a,d), gas phase with a 1 V homogeneous field (panels b,e), and gas phase with a 2 V field (panels c,f); panels g–i show the analogous sequence with implicit aqueous solvation. Color bar range shown in the right. Warm colors indicate regions of relative positive potential , cool colors indicate relative negative potential .

Figure 5 illustrates the molecular electrostatic potential (MESP) mapped onto the electron density isosurfaces for neutral saccharin and the saccharinate anion. The maps are compared under three conditions:  in the gas phase without the presence of any external electric field, in the gas phase under the influence of 0.013 a.u. and 0.026 a.u. electric fields that correspond to a 1V and 2V electric potential drop, respectively. Panels (a–f) present the gas-phase results, while panels (g–i) show the analogous systems with an implicit aqueous solvation model. The color scheme denotes the electrostatic character. Warm colors (red/orange) indicate regions of positive potential (electrophilic), while cool colors (blue) indicate regions of negative potential (nucleophilic). In all cases, electron rich regions are localized on the sulfonyl oxygens and heteroatoms of the heterocyclic ring , whereas electron deficient regions occur near the hydrogen bearing portions of the molecule. Critically, the saccharinate anion exhibits a stronger and more delocalized negative potential, particularly around the deprotonated ring nitrogen and the sulfonyl oxygens, which is a direct result of the extra electron density upon deprotonation.

The addition of implicit aqueous solvation(g-i) represents a key competitor. This inclusion raises an important question about a solvent cavity, which has a profound effect of reducing the field polarization effects seen in gas-phase calculations. The molecular electrostatic potential surfaces of both species in solution are also found to be nearly symmetric, with a closer proximity to their zero-field gas-phase structures than their respective gas-phase, field-polarized geometries. This also provides a hint that the electrical field from the electrode's influence on adsorption orientation in plating experiments is likely opposed by a very fine balance between a randomizing, stabilizing solvent sheath influence[11].

The ESP distribution provides a molecular-level explanation for adsorption. The sulfonyl oxygens with their high nucleophilicity and ring nitrogen provide the major site for Lewis basicity, coordinating with electrophilic metal centers. This can involve either solvated $Ni^{2+}$ ions in the electrolyte or partially charged Ni atoms on the cathode surface, facilitating inner sphere adsorption. Furthermore, the applied electric field, which mimics the interfacial field at an electrode, demonstrably perturbs the ESP landscape. This field effect can polarize the molecule, thereby modulating both the strength and geometry of adsorption by enhancing electrostatic complementarity with the metal surface and aligning the molecular dipole. Therefore, this field-induced polarization is a key mechanistic factor, as it can "pre-orient" the molecule for adsorption. A molecule that might present a neutral or repulsive face to the surface in the absence of a field can be reoriented to present a highly attractive, nucleophilic, or electrophilic region when a cathodic or anodic potential is applied[25].

## DFT approach-energy analysis

To probe how solvation and interfacial electric fields bias saccharinate and saccharin adsorption geometries, we computed approach energies (ΔE) for seven approach vectors of them toward a Ni in gas and implicit aqueous phases, with and without an applied external field (Table 2,3). It should be note that since everything here was done with an implicit solvent model and uniform electric field, the adsorption energies shouldn't be taken as exact numbers. The data reveal two dominant trends: First, in the gas phase, the interactions are generally exothermic (ΔE ranging from −26 to −44 kcal mol$^{-1}$), indicating favorable adsorption of the saccharinate anion to the Ni surface, with the application of an external field further stabilizing specific orientations, especially in directions 5 and 6 (ΔE ≈ −50 kcal mol$^{-1}$), consistent with field induced molecular alignment. Gas phase calculations further indicate intrinsically strong O- and N-directed binding, but in solution the N-facing geometry is rendered only weakly favorable and is destabilized by the interfacial field.Therefore, although N-binding is thermodynamically favorable in isolated complexes, solvation and electrostatics can shift the preferred coordination mode at an electrode interface. This field-driven preference likely reflects dipole reorientation that maximizes donor/acceptor overlap while minimizing electrostatic penalty.

In aqueous solution and in the absence of a field, some orientations remain weakly favorable (ΔE ≈ −6 kcal mol$^{-1}$), while others, such as direction 3,4 and 7, become highly destabilized, reflecting the energetic penalty associated with desolvation. Under aqueous plus field conditions, most ΔE values become strongly positive, suggesting that the saccharinate anion is electrostatically repelled from the surface. However, one orientation (direction 7) remains exothermic (ΔE = −20.65 kcal mol$^{-1}$). This is consistent with the idea that, despite its strong solvation, adsorption of the anion can still occur at locally desolvated sites on the cathode. There are nanoscale domains where water is reduced due to either the high electric field in the double layer, solvent thermal fluctuations, or surface topography including steps, terraces, or defects. These also decrease the free energy cost of removing solvent. In addition, the high electric field in the inner Helmholtz plane can reorient solvent molecules, compress the solvation layer, or create pockets of reduced hydration. As a result, even well-solvated anions like saccharinate can reach the surface in specific orientations.

Table 2. Computed approach energies (ΔE, kcal·mol⁻¹) for the saccharinate anion at seven defined approach vectors toward nickel. Columns report energies in aqueous and gas phases, with and without an applied external field (x = 1 V).

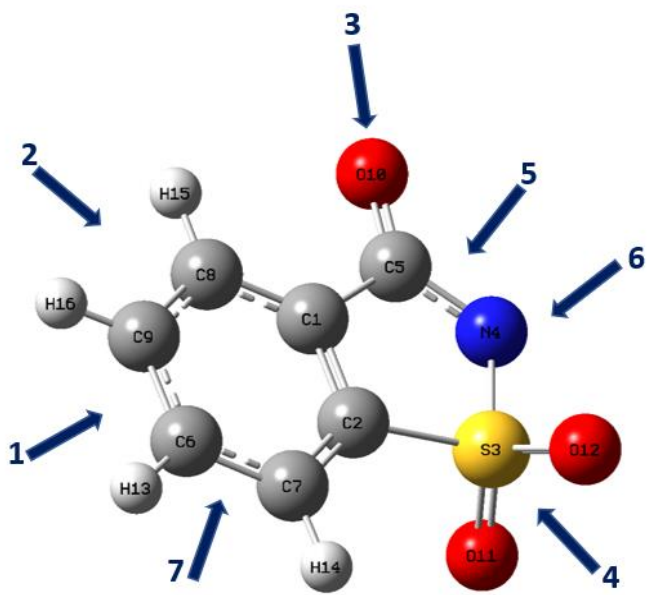

| Direction | ΔE (aq, field x = 1 V) (kcal·mol⁻¹) | ΔE (aq, no field) (kcal·mol⁻¹) | ΔE (gas, field x = 1 V) (kcal·mol⁻¹) | ΔE (gas, no field) (kcal·mol⁻¹) |
|---|---|---|---|---|
| 1 | 116.89 | –6.23 | –45.46 | –25.89622 |
| 2 | 84.37 | –6.26 | 25.83 | –27.75679 |
| 3 | 42.07 | 30.46 | –21.05 | –40.07774 |
| 4 | 56.09 | 8.09 | –23.63 | –32.82249 |
| 5 | 51.57 | –9.72 | –50.09 | –44.34851 |
| 6 | 15.45 | –9.70 | –50.00 | –44.34851 |
| 7 | –20.65 | 140.12 | 23.38 | –38.22948 |

Table 3 presents the corresponding ΔE values for neutral saccharin. Here, the behaviour contrasts sharply with that of the anion. In the aqueous phase without an external field, all seven orientations exhibit strong exothermic interactions (ΔE ranging from −13.61 to −54.78 kcal mol⁻¹), indicating that neutral saccharin can readily approach the metal surface without significant electrostatic repulsion. This suggests that neutral saccharin does not face the solvation energy penalty observed with the anionic form. Under aqueous plus field conditions, the adsorption of neutral saccharin remains favorable for specific orientations, most notably directions 2 (ΔE = - 43.81 kcal mol⁻¹) and 4 (ΔE = - 49.71 kcal mol⁻¹), highlighting the role of the applied field in promoting preferential alignment of the sulfonyl oxygen or aromatic π-face towards the Ni surface. These findings suggest that neutral saccharin has a lower effective desolvation cost and therefore may act as a more surface accessible species. However, given that the PCM model does not capture specific solvation or proton-coupled interfacial processes, this conclusion should be viewed as a mechanistic hypothesis that is consistent with spectroscopic observations rather than a definitive assignment.

Coupled analysis of DFT calculations with electrochemical insights suggests that saccharin affects nickel electrodeposition as a result of interfacial adsorption rather than complexation[10, 13]. The dominant form of saccharin that interacts with nickel in an aqueous environment appears to be the molecular form, in which several paths of approach remain exothermic in the absence of an external field, due to a diminished degree of desolvation with increased access to the inner Helmholtz layer.

This characteristic environment, in which the electrical effects of ionic screening by co existent cations like $K^+$ and $H^+$, as well as additional specific adsorption of anions like $Cl^-$ above the potential of zero charge, are especially important, allows saccharin to adsorb either weakly in a physisorption reaction or in a more bound fashion, possibly in a reaction involving inner sphere interactions with atoms like oxygen and/or nitrogen. The magnitude of such interactions would then influence which fragments of sulfur and carbon are incorporated into a growing nickel film, with weakly bound components being removed easily compared to those that are bound in a manner that can lead to entrapment in a growing nickel film. As a result, saccharin adsorption reduces electron concentration and blocks active sites for reduction, which in turn increases site energy and creates surface potential inhomogeneity. This promotes texture variation, reduces grain size, and produces a finer, brighter deposit despite local inhomogeneity. Electron depletion also diminishes d-state density and conductivity along grain boundaries, which in turn enhances charge-transfer resistance. These effects are shown in experimentally observed lower cathodic currents, shifted reduction peaks in cyclic voltammetry, increased charge-transfer resistance values, and lower low-frequency impedance in impedance spectroscopy [26].

Table 3 Computed approach energies ($\Delta E$, kcal·mol$^{-1}$) for neutral saccharin at seven approach vectors to Ni surface. Results are shown for aqueous and gas phases, both with and without an applied external field (x = 1 V). Direction indices refer t

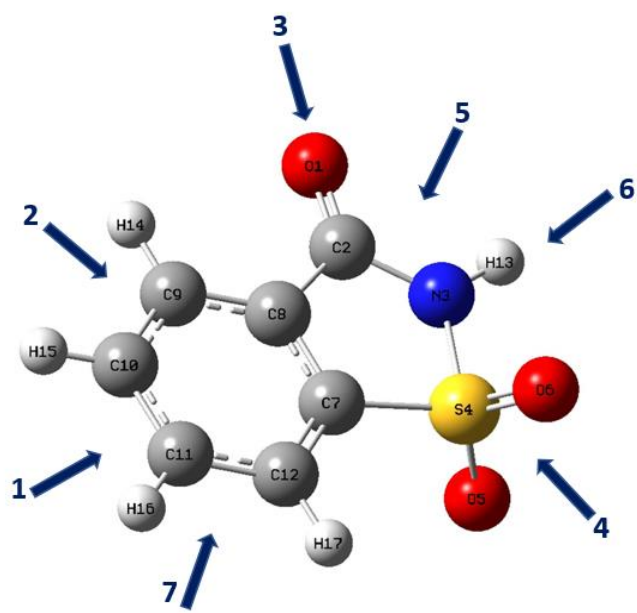

| Direction | $\Delta E$ (aq, field x = 1 V) (kcal·mol$^{-1}$) | $\Delta E$ (aq, no field) (kcal·mol$^{-1}$) | $\Delta E$ (gas, field x = 1 V) (kcal·mol$^{-1}$) | $\Delta E$ (gas, no field) (kcal·mol$^{-1}$) |
|---|---|---|---|---|
| 1 | 30.09 | –13.608 | 16.40 | –41.5618 |
| 2 | –43.81 | –48.054 | –9.22 | –41.1351 |
| 3 | 6.40 | –31.355 | –31.81 | –32.1421 |
| 4 | –49.71 | –50.055 | –23.98 | –27.9957 |
| 5 | 6.40 | –48.225 | –31.82 | –33.1739 |
| 6 | 1.33 | –54.775 | –30.24 | –39.5824 |
| 7 | 43.87 | –49.293 | –9.83 | –41.6923 |

Taken together, these orientation and field-dependent energetic trends are consistent with the potential dependent SERS signatures and support a mechanistic picture in which protonation state, local desolvation, field orientation, and competition with specifically adsorbed anions jointly determine the adsorption geometries that control saccharin's influence on Ni growth[8].Although the implicit-solvent approach does not include explicit solvent dynamics, ion pairing, or full periodic surface relaxation and therefore does not yield absolute adsorption free energies, it reliably captures the relative, direction-dependent trends; accordingly, the ΔE values should be interpreted qualitatively and comparatively when linking the computations to spectroscopic and electrochemical observations.Where possible, these qualitative predictions can be tested experimentally via potential dependent SERS, EQCM, or in situ STM studies that probe orientation, coverage, and local desolvation at the electrode interface.

## Conclusion

This density functional theory study elucidates the molecular scale adsorption mechanisms of saccharin and saccharinate on nickel under an electroplating-relevant interfacial electric field. By integrating frontier orbital analysis, molecular electrostatic potentials, and approach energies, we establish that adsorption is governed by a synergistic interplay of four key factors: (1) molecular protonation state, (2) solvation and desolvation energetics, (3) field induced polarisation and dipole alignment, and (4) local surface topology. A critical observation from the calculations is that the saccharinate anion remains strongly stabilized by its solvation environment. Although PCM does not explicitly capture solvent/solute hydrogen bonding or dynamic desolvation, the relative trends in the computed approach energies qualitatively indicate a higher desolvation penalty for the anion, which reduces its likelihood of approaching the electrode surface. In contrast, neutral saccharin, though a minor species in solution, adsorbs more readily due to a lower solvation barrier. The applied electrical field does not act uniformly but selectively stabilizes specific adsorption geometries that expose donor-rich O/N sites or enable favorable π-metal overlap , while destabilizing others. This field selective stereochemical control, combined with transient local phenomena like solvent fluctuations and adsorption at defects, reconciles the apparent contradiction between bulk anionic speciation and the observed surface activity of saccharin. These insights provide a mechanistic link between molecular behavior and macroscopic plating outcomes. Also, present a scientific basis for moving beyond trial and error and toward the rational design of more effective plating additives.